\documentclass[prl,aps,twocolumn,nofootinbib]{revtex4}

\usepackage{bm}
\usepackage{subfigure}
\usepackage{amsmath,amssymb}
\usepackage{color,multirow}
\usepackage{graphicx}
\usepackage{epstopdf}
\usepackage[samesize]{cancel}

\usepackage{bm}

\begin{document}
\title{Supersymmetry in the Insulating phase of a Chain of Majorana Cooper Pair boxes}

\author{Hiromi Ebisu, Eran Sagi, and Yuval Oreg
}
\affiliation{
Department of Condensed Matter Physics, Weizmann Institute of Science, Rehovot, Israel 76100\\
}
\date{\today}
\begin{abstract}
The charging energy $U$ of a small superconducting island containing Majorana zero modes---a Majorana Cooper-pair box---induces interactions between the Majorana zero modes. Considering a chain of many such boxes, a topological superconductor-insulator transition occurs when $U$ is much larger than the transfer matrix element $t$ between the boxes.
In this Letter, we focus on the insulting phases occurring in this regime. We show that there are several competing insulating phases, and that the transition between them is described by a supersymmetric field theory with a central charge $c=7/10$. We obtain this result by mapping the model to a spin-$1$ system and through a field theoretical approach.  The microscopic model we propose consists of a chain of Majorana Cooper-pair boxes with  \textit{local} tunneling between Majorana zero modes and \textit{local} charging energy terms, which can be controlled by gate potentials, thus making its realization more feasible.
\end{abstract}
\pacs{pacs}
\maketitle

\thispagestyle{empty}
\textit{Introduction.}---Majorana zero modes (MZMs) consisting of an equal superposition of electrons and holes have been tremendously important in condensed matter physics. They shed light on new aspects of physics, such as
topological phases of matter \cite{Kane2010} and non-Abelian statistics \cite{Nayak2008}. In recent years, considerable theoretical and experimental efforts have been devoted to realizations of MZMs on the ends of topological nanowires \cite{Oreg2010,Lutchyn2010,Sau2010,Deng2012,Rokhinson2012,Mourik2012,Vaitiekenas2018}.
Remarkably, signatures of MZMs have been reported in experiments exhibiting charging effect~\cite{Albrecht2016} (for a recent review see Ref. \cite{Lutchyn2018}). \par

The conventional Cooper-pair box, a small superconducting island with charging energy $U$, has been intensively studied as a playground for studying the fundamental superconductor-insulator transition (for a review see Ref. \cite{Schon2001}). For example, Ref. \cite{Glazman1997} uncovered that if the gate potential on each box is set to a specific value, the system has twofold degeneracy, leading to a repulsive Luttinger liquid phase in addition to the insulator and superconducting phases. As we discuss here, a Majorana Cooper pair box (MCB)~\cite{hassler2012strongly,Barkeshli2015,Plugge2016,plugge2017majorana,karzig2017scalable}, has additional degrees of freedom, i.e., the MZMs, which may lead to an even richer phase diagram.

In Ref. \cite{spinliquid}, a MCB hosting six MZMs (the so-called hexon) was proposed as a basic element for constructing a one-dimensional insulating phase that is equivalent to the transverse field Ising spin chain, and the two-dimensional insulating Kitaev's spin liquid phase~\cite{Kitaev2006} utilizing the Yao-Kivelson model \cite{Yao2007}. Furthermore, by tuning various parameters, it was possible to show that in the insulting state a transverse field Ising phase transition, described by a $1+1$-dimensional conformal field theory \cite{Belavin1984} (CFT) with central charge $c=1/2$, is stabilized.

The purpose of this Letter is to go beyond the work presented in Ref. \cite{spinliquid}, and demonstrate that a quantum critical point, described by a supersymetric (SUSY) conformal field theory with central charge $c=7/10$, may occur in the insulating phase by properly tuning the system parameters.\par
\begin{figure}[htbp]
	\subfigure[]{%
		\includegraphics[clip, width=0.3\columnwidth]{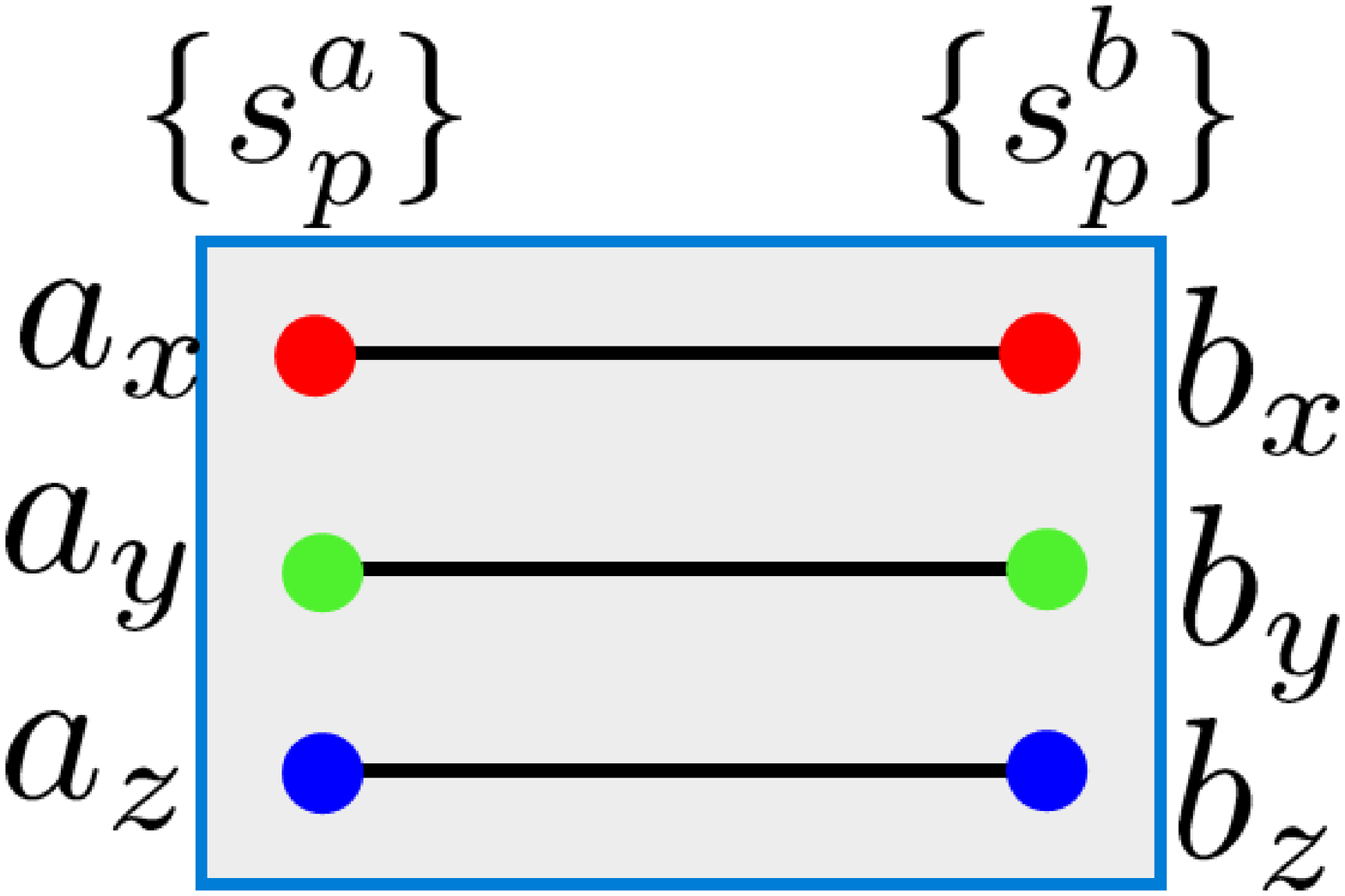}}%
	\subfigure[]{%
		\includegraphics[clip, width=0.6\columnwidth]{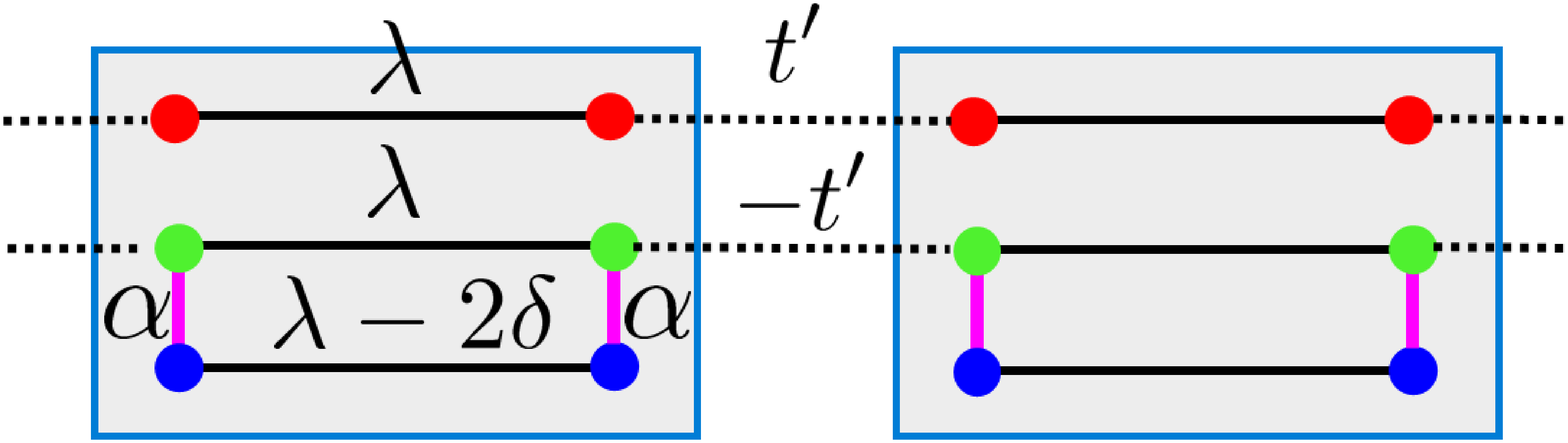}}%
	\caption{(a) The hexon---a Majorana Cooper-pair box hosting six Majorana zero modes. The gray box represents a superconducting island, with charging energy $U$, and the black solid lines depict topological nanowires with MZMs,  portrayed by solid dots, on each end. 
In accordance with the index $p=x,y,z$, the MZMs are painted in different colors. In the strong charging energy regime, the  $a_p$ ($b_p$) triplet of MZMs forms a spin-$1/2$ degrees of freedom denoted by $s^a_p$ ($s^b_p$).  (b) The configuration of the couplings of the Hamiltonian $H_0$ [Eq. (\ref{hami2})] which constructs the Blume-Capel model. }
	\label{fg1}
\end{figure}

We use two distinct approaches to demonstrate the emergence of the SUSY in the insulating phase, when the charging energy on the box $U$ is much larger than the hopping between the boxes $t'$, cf. Fig.~\ref{fg1}. In the first approach, we explicitly construct the Blume-Capel~\cite{Blume,Capel} (BC) spin-$1$ model from a chain of MCBs, which is known numerically to exhibit emergent SUSY~\cite{Mussardo2017}. In the second approach, we consider an effective field theory by focusing on the low-energy continuum limit. The latter will be shown analytically to be described by a super Landau-Ginzburg (LG) action which exhibits the same universality as the superconformal field theory (SCFT) with $c=7/10$ \cite{Zamolodchikov1986}. The two approaches we present complement each other, with one relying on numerical results, while the other consists entirely of analytical considerations.
\par
Previous works have already discussed the possibility of emergent SUSY in interacting MZM systems~\cite{Kastor1989, Affleck2015,Franz2016,Grover2014,o2018lattice}. However, the model discussed here is based on a concrete microscopic model, requiring only \textit{local} couplings of MZMs and charging energy, which can be controlled by gates and therefore may be a more feasible proposal. \par
\textit{The Majorana Cooper-pair Box (MCB) chain}.---We consider placing three semiconducting nanowires, labeled by the index $p=x,y,z$, on top of the superconducting island, see Fig. \ref{fg1}(a).  Assuming these nanowires are proximity coupled to the superconducting island, under application of a  magnetic field parallel to the wires, each nanowire hosts
two MZMs on its ends. These two MZMs can accommodate a fermion. Introducing the six Majorana operators $a_p$, $b_p$, obeying the anti-commutation relations $\{a_{p},a_{q}\}=\{b_{p},b_{q}\}=2\delta_{p,q}$ \cite{Oreg2010,Lutchyn2010,Sau2010}, we denote the annihilation operator of these fermions by $\frac{1}{2}\left(a_{p}+i b_{p}\right)$. For each pair of MZMs, the $\mathbb{Z}_2$ fermion parity takes the values $ia_{p}b_{p}=\pm1$. \par
We describe a system consisting of a one-dimensional array of MCBs with six MZMs each using the Hamiltonian $H=H_U+H_0$, 
with $H_U$ being the charging energy term and $H_0$ representing the 
general hopping elements between MZMs:  
\begin{eqnarray}
H_U&=&\sum_{j}U(2\hat{N}_c^j+{n}_g^j+\hat{n}_M^j)^2,\label{hami}\\
H_0&=&i\sum_{j,j^{\prime},p}t_p^{jj^{\prime}}a_p^{j}b_p^{j^{\prime}}+i\sum_{j,p,p^{\prime}}h^{aj}_{pp^{\prime}}a_p^ja^j_{p^{\prime}}+h^{bj}_{pp^{\prime}}b_p^jb^j_{p^{\prime}}\label{hami2}.
\end{eqnarray}
Here, the superscript $j$ labels the $j$th MCB, $U$ is the charging energy of the box, ${n}_g^j$ is a continuous value controlled by a gate potential, $\hat N_c^j$ is the number operator of Cooper-pairs in the box, and $\hat n_M^j=\sum_{p}(1-ia_{p}^jb^j_{p})/2$ is the number of fermions occupying the MZMs.

\par
\textit{Construction of the Blume-Capel (BC)  model}.---Before describing our first approach, let us briefly review the BC model \cite{Blume,Capel}.

By a-classical-to-quantum mapping \cite{Fradkin1978}, the two-dimensional classical BC model is mapped onto a one-dimensional quantum model, given by the Hamiltonian
\begin{equation}
H_{\text{BC}}=\sum_j\alpha S^j_x+\delta (S^j_z)^2-JS^j_zS^{j+1}_z\label{bc},
\end{equation}
where $S^j_{z,(x)}$ are spin-$1$ operators along the $z$ ($x$) axis  at site $j$.
The phase diagram is special as it has a first order transition line that meets a second order line at a tricritical fixed point.
At the tricritical fixed point, a CFT with $c=7/10$ emerges \cite{Mussardo2017}. This CFT is further known to possess $\mathcal{N}=1$ SUSY \cite{Friedan}.
Indeed, numerical studies confirmed the existence of a tricritical fixed point for finite  $\alpha$ and  $\delta$ \cite{Saul,Gehlen}. \par
We now present the first approach to obtain the emergent SUSY, in which we construct the BC model given in Eq.~(\ref{bc}). We begin with the Hamiltonian $H=H_U+H_0$ defined in Eqs. (\ref{hami}) and (\ref{hami2}), and set the (real) couplings in $H_0$ as $t^{jj}_x=t^{jj}_y=\lambda$, $t^{jj}_z=\lambda-2\delta$, $h^{aj}_{yz}=h^{bj}_{yz}=\alpha$, $t^{jj+1}_x=-t^{jj+1}_y=t^{\prime}$, with all other couplings set to zero [See Fig.~\ref{fg1}(b)].
Below, we will see how this Hamiltonian reproduces the BC model when $U\gg t^{jj^{\prime}}_p, h_{pp^{\prime}}^{a/bj}$. \par
The six MZMs in each hexon define spin-$1/2$ operators \cite{spinliquid}
\begin{eqnarray}
s^{a^j}_x&=&ia^j_ya^j_z, \;s^{aj}_y=ia^j_xa^j_z,\; s^{aj}_z=ia^j_xa^j_y\nonumber\\
s^{b^j}_x&=&ib^j_yb^j_z, \;s^{bj}_y=ib^j_xb^j_z,\; s^{bj}_z=ib^j_xb^j_y.\label{spin}
\end{eqnarray}
It is straightforward to check that Eq.~(\ref{spin}) satisfies the spin-$1/2$ algebra. That is, $(s^{a^j}_{p})^2=1$, $[s^{a^j}_p,s^{a^j}_q]= i \epsilon_{pqk} s^{a^j}_k$ (and similarly for $\{s_{p}^{b^j}\}$), with $\epsilon_{pqk}$ being the anti-symmetric tensor. 
Because of the strong charging energy $U$ the number of pairs in each MCB is fixed, leading to a constraint on the total $\mathbb{Z}_2$ fermion parity of each hexon. Such constraint reads as \cite{spl}
\begin{equation}
(ia^j_xb^j_x)(ia^j_yb^j_y)(ia^j_zb^j_z)=1\label{parity}.
\end{equation}
This constraint ensures that the total number of states in each hexon is four, which is identical to the number of states of two spin-$1/2$ degrees of freedom.
A key step of our construction is to project out the singlet state of the total spin  ${S}^{j}_{p}=s^{a_j}_{p}+s^{b_j}_{p}$ of each MCB. This allows us to obtain spin-$1$ (spin-triplet) states. Such a projection can be implemented by introducing couplings of the MZMs as
\begin{equation}
H_{\lambda}=i\lambda\sum_{j,p=x,y,z }	a^j_{p}b^j_{p}\label{Hl}
\end{equation}
and setting $\lambda>0$. To see this, we note that the norm of the total spin is written as $\bm{S}^j\cdot\bm{S}^j= 2+2\bm{s}^{a^j}\cdot\bm{s}^{b^j}$; thus using Eq. (\ref{spin}) and the constraint in Eq.~(\ref{parity}), we find
\begin{equation*}
\bm{s}^{a^j}\cdot\bm{s}^{b^j} = -i\sum_pa^j_pb^j_p.
\end{equation*}
Therefore, we have $H_{\lambda}= -\frac{\lambda}{2}\sum_j\bm{S}^j\cdot\bm{S}^j+\lambda$, with $\lambda>0$. 
Increasing $\lambda$, the spin-$1$ (spin-triplet) states become energetically favored. Thus, focusing on low energies, the spin-singlet state is projected out. Below, we use $\bm{S}^{j}$ to denote spin-$1$ operators. \par
%

Having a spin-$1$ degree of freedom on each box, we can reproduce the BC model. The first term of the BC model is obtained by
\begin{equation}
H_{\alpha}=i\alpha\sum_j(a^j_ya^j_z+b^j_yb^j_z)=\sum_j\alpha S_x^j.\label{ha}
\end{equation}
Indeed, $ia^j_ya^j_z+ib^j_yb^j_z=s^{a^j}_x+s^{b^j}_x=S^j_x$. The second term of the BC model, $\delta(S^j_z)^2$ is realized by
\begin{equation}
H_{\delta}=-2\delta\sum_jia^j_zb^j_z=\sum_j\delta(S_z^j)^2-2\delta,\label{hd}
\end{equation}
as $\frac{1}{2}(S^j_z)^2-1=s^{a^j}_zs^{b^j}_z=(ia^j_xa^j_y)(ib^j_xb^j_y)=-ia^j_zb^j_z$, where the last equality follows from Eq. (\ref{parity}). \par
To reproduce the last term of the BC model, $S^j_zS^{j+1}_z$, we consider small magnitude of couplings of MZMs between adjacent islands:
\begin{equation*}
i\sum_{j}t^{\prime}(b^j_xa^{j+1}_x-b^j_ya^{j+1}_y).
\end{equation*}
Because of the strong charging energy,
these couplings are regarded as a perturbation. Using a Schrieffer-Wolff transformation \cite{Wolf}, we obtain the following Hamiltonian, which is further mapped to the last term of the BC model after the projection to the spin-$1$ states:
\begin{eqnarray}
H_J&=&J\sum_j(ib^j_xa^{j+1}_x)(ib^j_ya^{j+1}_y)\nonumber\\
   &=&-J\sum_j s^{b^j}_zs^{a^{j+1}}_z\simeq \sum_j-JS^j_zS^{j+1}_z,\label{hj}
\end{eqnarray}
where $J={t^{\prime}}^2/2U$. 
For a derivation of the last relation, see the Supplemental Material~\cite{spl}.
\par
The terms in Eqs. (\ref{ha}), (\ref{hd}), and (\ref{hj}) establish the mapping between the chain of MCBs and the BC spin-$1$ model.
It was found numerically \cite{Gehlen} that for $\alpha/J \simeq 0.9$ and $\delta/J \simeq 0.4$, the phase of the MCBs reaches the tricritical fixed point. At this point, SUSY emerges and the long distance behavior of the BC model is characterized by SCFT with central charge $c=7/10$~\cite{Friedan}.
\par

\textit{Field theoretical approach.}---We move on to the second field theoretical approach.

\begin{figure}[htbp]
 \begin{center}
  \includegraphics[width=50mm]{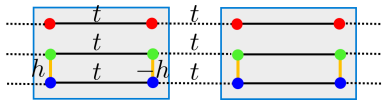}
 \end{center}
 \caption{The configuration of the MCB chain corresponding to the Hamiltonian given in Eqs. (\ref{ht}) and (\ref{hh}). }
 \label{fg2}
\end{figure}

Similarly to the first approach, we assume that the charging energy is larger than a typical coupling between the MZMs. We start with the model Hamiltonian in Eqs.~(\ref{hami}) and (\ref{hami2}), which was mapped in the previous section to the BC model. However, to facilitate the field theoretical analysis, we choose the parameters in  $H_0$ to be $t^{jj}_p=t^{jj+1}_p=t$, $h^{aj}_{yz}=-h^{bj}_{yz}=h$ (see also Fig. \ref{fg2}). In this configuration, $H_0$ can be rewritten as $H_0=H_t+H_h$ with
\begin{eqnarray}
H_t&=&it\sum_{j,\substack{p=x,y,z}}b^j_{p}(a^{j+1}_{p}-a^j_{p}),\label{ht}\\
H_h&=&ih\sum_j(a^j_ya^j_z-b^j_yb^j_z).\label{hh}
\end{eqnarray}	
The Hamiltonian in Eq.~(\ref{ht}) describes three critical ``Majorana chains'' yielding three left and right moving Majorana fields at low energies. To obtain a continuum low energy description,
we use the standard approach~\cite{Gogolin} and replace the MZM operators $a^j_{p}$ and $b^j_{p}$ with slow varying Majorana fields
$a_{p}^j \simeq \alpha_{p}(r) , b^j_{p}\simeq\beta_{p}(r)$ (with $r$ being the coordinate along the array of MCBs). The Hamiltonian density from Eq.~(\ref{ht}) is then modified to 
\begin{equation*}
\mathcal{H}_t\simeq i\sum_{p}t\beta_{p}(r)\partial_r\alpha_{p}(r).\label{cf}
\end{equation*}
Transforming the fields to a ``chiral basis'' via
\begin{equation*}
\alpha_{p}(r)=\frac{-\eta_{pR}+\eta_{pL}}{\sqrt{2}},\;\;
\beta_{p}(r)=\frac{\eta_{pR}+\eta_{pL}}{\sqrt{2}},
\end{equation*}
where $\eta_{pR/L}$ denotes the right-left moving Majorana field,
$\mathcal{H}_t$ and the Hamiltonian density from Eq. (\ref{hh}), $\mathcal{H}_h$ becomes
\begin{eqnarray*}
\mathcal{H}_t&=&\frac{i}{2}\sum_{p}t(\eta_{pR}\partial_r\eta_{pR}-\eta_{pL}\partial_r\eta_{pL}),\\
\mathcal{H}_h&=&-ih(\eta_{yR}\eta_{zL}+\eta_{yL}\eta_{zR}).
\end{eqnarray*}
Defining the Majorana spinor $\psi_{p}=(\eta_{pR},\eta_{pL})^T$ and the Dirac gamma matrices $\gamma_0=\sigma_y, \gamma_1= i\sigma_x$ (where $\sigma_{x,y}$ denotes the $2\times 2$ Pauli matrices), we get the $1+1$-dimensional Lagrangian density\cite{foot1}
\begin{equation}
\mathcal{L}=\sum_{p}\frac{t}{2}\bar{\psi}_{p}i\cancel \partial\psi_{p}+ih(\eta_{yR}\eta_{zL}+\eta_{yL}\eta_{zR}),\label{leff}
\end{equation}
with $\bar{\psi}_p=\psi_p^{T}\gamma_0$,  $\cancel \partial=\partial_{\mu}\gamma^{\mu}$.

For $U \gg t$ we can perform the Villain approximation, and treat $t$ perturbatively, yielding the  interacting term
\begin{equation}
-g \cos\Bigl(\frac{\pi}{4}\sum_p\bar{\psi}_p\psi_p\Bigr)\label{int}
\end{equation}
in the Lagrangian, where $g$ relates to $U$ in Eq.~(\ref{hami}) by $g\simeq \frac{U}{\pi^2}$. The details of the derivation are described in the Supplemental Material \cite{spl}.
To analyze the effect of the terms proportional to $h$ and $g$, we implement the bosonization procedure. To do so, we form a fermion out of $\eta_{yR/L}$ and $\eta_{zR/L}$, which is then bosonized:
\begin{equation*}
\Psi_{R/L}=\eta_{yR/L}+i\eta_{zR/L}\simeq e^{\pm i\sqrt{4\pi}\phi_{R/L}},
\end{equation*}
where $\Psi_{R/L}$ and $\phi_{R/L}$ indicate right-left moving Dirac and boson fields, respectively. An important consequence of the bosonized formulation is that we obtain one bosonic field and one Majorana  (real fermion) field . This hints at the possibility of SUSY, where the number of bosonic degrees of freedom is equal to the fermionic one.
Denoting $\psi_x \rightarrow \psi$, we find the Lagrangian density
\begin{eqnarray}
\mathcal{L}&=&\frac{1}{2}(\partial_{\mu}\varphi)^2+\frac{1}{2}\bar{\psi}i\cancel \partial\psi \nonumber\\
&-&g(\sin^2\sqrt{4\pi}\varphi-2\cos\sqrt{4\pi}\varphi\bar{\psi}\psi)
\notag\\
&+&h\sin\sqrt{4\pi}\varphi,\label{leff2}
\end{eqnarray}
with $\varphi=\phi_R+\phi_L$. Here, we have rescaled the fields and normalized $t$ to be unity.
\par

To complete our analysis, we assume further that $h$ and $g$ are positive and $h > g > 1$. Focusing on low energies, we can thus expand the boson field around the minimum of the Hamiltonian (or the maximum of the Lagrangian), $\varphi\simeq\frac{\pi}{2}\frac{1}{\sqrt{4\pi}}+\tilde{\varphi}$, resulting in the expanded Lagrangian density
\begin{eqnarray}
\mathcal{L}&=&\frac{1}{2}(\partial_{\mu}\tilde{\varphi})^2+\frac{1}{2}\bar{\psi}i\cancel \partial\psi\\
&+&2g(-\sqrt{4\pi}\tilde{\varphi})\bar{\psi}\psi-g\Bigl(1-\frac{4\pi}{2}\tilde{\varphi}^2\Bigr)^2\\
&+&h\Bigl(1-\frac{4\pi}{2}\tilde{\varphi}^2\Bigr)\label{lp}.
\end{eqnarray}
Tuning $g=\pi/2$ and $h=\pi$, ${\cal L}$ is further simplified to
\begin{equation}
\mathcal{L}\simeq
\frac{1}{2}(\partial_{\mu}\tilde{\varphi})^2+\frac{1}{2}\bar{\psi}i\cancel \partial\psi-\frac{1}{2}v\tilde{\varphi}\bar{\psi}\psi-\frac{1}{8}v^2\tilde{\varphi}^4\label{susy},
\end{equation}
where $v=2\pi\sqrt{4\pi}$. Remarkably,
Eq.~(\ref{susy}) is identical to the $\mathcal{N}=1$ super LG action. The relation to the super LG action can be obtained explicitly by considering the SUSY model
\begin{equation}
S_{\text{SUSY}}=\int dxdt\;d\theta^2\Bigl[\frac{1}{4}(\bar{D}\Phi)(D\Phi)+W(\Phi)\Bigr],\label{susy2}
\end{equation}
where $\Phi$ is the superfield defined by $\Phi=\tilde\varphi+\bar{\theta}\psi+\frac{1}{2}\bar{\theta}\theta F$, $D$ represents covariant derivative in superspace, and $W(\Phi)$ describes superpotential which is a polynomial function of~$\Phi$~\cite{DIVECCHIA}. In our case, $W(\Phi)$ is given by
$W(\Phi)=\frac{v}{6}\Phi^3$~ \cite{foot3}. 
Referecence \cite{Zamolodchikov1986} shows that at long distances, the super LG action with a superpotential $W(\Phi)\simeq \Phi^m\;(m=2,3,\cdots)$ exhibits a supersymmetric analog of the minimal models, characterized by central charge $c=\frac{3}{2}-\frac{12}{m(m+2)}$. Since our case corresponds to $m=3$, the theory given in Eq. (\ref{susy}) effectively manifests an emergent SUSY described by a SCFT with $c=7/10$. \par
While we chose $g=\pi/2$ and $h=\pi$ above, we can instead follow Ref.~\cite{Witten} and realize an identical SCFT for generic values of $g$ by tuning $h$ properly. Indeed, redefining $\tilde{\sigma}=\sqrt{K}\tilde{\varphi}$, $u=4g\sqrt{\frac{4\pi}{K}}$, $K=1-\frac{4g}{\pi}\rho$, $\rho=\frac{2g/\pi-1}{8g^2/\pi^2-1}$, $h=2(1-\rho)g$, the theory in Eq.~(\ref{lp}) takes the form
\begin{equation}
\mathcal{L}\simeq
\frac{1}{2}(\partial_{\mu}\tilde{\sigma})^2+\frac{1}{2}\bar{\psi}i\cancel \partial\psi-\frac{1}{2}u\tilde{\sigma}\bar{\psi}\psi-\frac{1}{8}u^2\tilde{\sigma}^4, \label{susy3}
\end{equation}
which is again equivalent to the super LG theory with the superpotential $W(\Phi)=u\frac{\Phi^3}{6}$\cite{foot3}. Notice that the expression for $\rho$ requires $g/t> \pi/(2 \sqrt{2})$ and that we can tune $h>g\gg 1$, which is consistent with our initial assumption that $U\gg t$.

\par
\begin{figure}[htbp]
 \begin{center}
  \includegraphics[width=30mm]{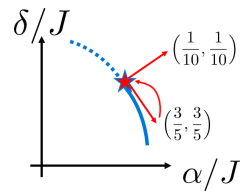}
 \end{center}
 \caption{A schematic picture of the phase diagram of the BC model. The red star depicts the tricritical fixed point and the solid (dashed) line represents the first (second) order transition line. Two red arrows indicate
perturbations in the tangential and orthogonal directions to the phase transition lines, whose conformal dimensions are given by $(\frac{3}{5},\frac{3}{5})$ and $(\frac{1}{10},\frac{1}{10})$, respectively. }
 \label{phase}
\end{figure}
\textit{Stability of the emergent SUSY to local perturbations.}---We can analyze the stability of the emergent SUSY using knowledge of the operator content of the SCFT. We focus on the case of the BC model. Suppose the BC model (constructed from the MCBs) is tuned to the tricritical fixed point, i.e., $(\alpha,\delta)=(\alpha_c,\delta_c)$ ($\alpha_c\simeq0.9$, $\delta_c\simeq0.4$ with $J$ being unity). We introduce a small deviation of $(\alpha,\delta)$ from the critical value, $(\alpha,\delta)\to(\alpha+\kappa_{\alpha},\delta+\kappa_{\delta})$ at specific site of the MCBs, $j=j_0$. Here, $\kappa_{\alpha/\delta}$ represents an infinitesimal deviation. Consider shifting the parameters $(\alpha,\delta)$ tangentially (orthogonally) to the phase transition line at the tricritical fixed point; see Fig. \ref{phase}.
Such a deviation can be done through a linear combination of $S_{x}^{j_0}$ and $(S_z^{j_0})^2$, as these two terms are realized by local couplings of MZMs in the MCBs.\par
At low energies, this situation can be described by a $c=7/10$ CFT which is perturbed by its primary fields. 
Moreover, the deviation of the parameters in the tangential (orthogonal) direction yields a perturbation given by a product of holomorphic and antiholomorphic primary fields of the form $\varepsilon_R\varepsilon_L[\varepsilon^{\prime}_R\varepsilon^{\prime}_L]$, with conformal dimension ($\frac{3}{5},\frac{3}{5}$)[($\frac{1}{10},\frac{1}{10}$)]~\cite{Kastor1989,Christe1990}. \par
Our consideration here is reminiscent of the localization problem of a single local impurity in a Luttinger liquid \cite{KaneFisher}. Similarly to this problem, we can judge whether the perturbation is relevant or not by following renormalization group equation
\begin{equation*}
\frac{d\mathcal{O}}{dl}=(1-\Delta_{\mathcal{O}})\mathcal{O},
\end{equation*}
where $\mathcal{O}$ is either $\varepsilon_R\varepsilon_L$ or $\varepsilon^{\prime}_R\varepsilon^{\prime}_L$, $l$ represents the logarithmic
rescaling factor, and $\Delta_O$ is the scaling dimension of $\mathcal{O}$.
Since the scaling dimension of $\varepsilon_R\varepsilon_L(\varepsilon^{\prime}_R\varepsilon^{\prime}_L)$
is $\frac{6}{5}(\frac{1}{5})$, the $c=7/10$ fixed point, and thus the emergent SUSY, is robust (sensitive) with respect to the tangential (orthogonal) perturbation. \par
In summary, we have studied the insulating phase of a chain of Majorana Cooper-pair boxes each of which has six Majorana zero modes, forming an ``hexon''. We include only local charging energy and local tunneling matrix elements in the model and identified a SUSY quantum phase transition with central charge $c=7/10$. This result was demonstrated by mapping
the system onto a spin-$1$ (the Blume-Capel) model and a field theoretical model.
To detect an imprint of the $c=7/10$ criticality, the thermal conductance would be a useful observable, similar to recent experiments in non-Abelian Hall systems~\cite{banerjee2018observation}. \par

It would be interesting to extend our considerations to the two-dimensional case through a wire construction. One can expect a two-dimensional topological phase with a chiral edge mode carrying central charge $c=7/10$.
Such a phase supports universal quantum computation~\cite{Nayak2008}.
This analysis is left to future projects.
\par
We thank A. Altland, N. Andrei, E. Berg, R. Egger, K. Flensberg, A. Keselman, R. Lutchyn, D. Pikulin, A. Stern, and Y. Tanaka for discussion.
This work was partially supported by the Israeli Science Foundation (ISF), the
Deutsche Forschungsgemeinschaft (CRC 183), the Binational Science
Foundation (BSF), the European Research Council under the European Union’s Seventh Framework
Programme (FP7/2007-2013)/ERC Grant Agreement, MUNATOP No. 340210, the European Union’s Horizon 2020 research and innovation programme (Grant Agreement LEGOTOP No 788715), the Japan Society for the Promotion of Science overseas
research fellowship, and The Adams Fellowship Program of the Israel Academy of Sciences and Humanities.

\bibliographystyle{apsrev-nourl}
\bibliography{ref}

\begin{thebibliography}{47}
\expandafter\ifx\csname natexlab\endcsname\relax\def\natexlab#1{#1}\fi
\expandafter\ifx\csname bibnamefont\endcsname\relax
  \def\bibnamefont#1{#1}\fi
\expandafter\ifx\csname bibfnamefont\endcsname\relax
  \def\bibfnamefont#1{#1}\fi
\expandafter\ifx\csname citenamefont\endcsname\relax
  \def\citenamefont#1{#1}\fi
\expandafter\ifx\csname url\endcsname\relax
  \def\url#1{\texttt{#1}}\fi
\expandafter\ifx\csname urlprefix\endcsname\relax\def\urlprefix{URL }\fi
\providecommand{\bibinfo}[2]{#2}
\providecommand{\eprint}[2][]{\url{#2}}

\bibitem[{\citenamefont{Hasan and Kane}(2010)}]{Kane2010}
\bibinfo{author}{\bibfnamefont{M.~Z.} \bibnamefont{Hasan}} \bibnamefont{and}
  \bibinfo{author}{\bibfnamefont{C.~L.} \bibnamefont{Kane}},
  \bibinfo{journal}{Rev. Mod. Phys.} \textbf{\bibinfo{volume}{82}},
  \bibinfo{pages}{3045} (\bibinfo{year}{2010}).

\bibitem[{\citenamefont{Nayak et~al.}(2008)\citenamefont{Nayak, Simon, Stern,
  Freedman, and Das~Sarma}}]{Nayak2008}
\bibinfo{author}{\bibfnamefont{C.}~\bibnamefont{Nayak}},
  \bibinfo{author}{\bibfnamefont{S.~H.} \bibnamefont{Simon}},
  \bibinfo{author}{\bibfnamefont{A.}~\bibnamefont{Stern}},
  \bibinfo{author}{\bibfnamefont{M.}~\bibnamefont{Freedman}}, \bibnamefont{and}
  \bibinfo{author}{\bibfnamefont{S.}~\bibnamefont{Das~Sarma}},
  \bibinfo{journal}{Rev. Mod. Phys.} \textbf{\bibinfo{volume}{80}},
  \bibinfo{pages}{1083} (\bibinfo{year}{2008}).

\bibitem[{\citenamefont{Oreg et~al.}(2010)\citenamefont{Oreg, Refael, and von
  Oppen}}]{Oreg2010}
\bibinfo{author}{\bibfnamefont{Y.}~\bibnamefont{Oreg}},
  \bibinfo{author}{\bibfnamefont{G.}~\bibnamefont{Refael}}, \bibnamefont{and}
  \bibinfo{author}{\bibfnamefont{F.}~\bibnamefont{von Oppen}},
  \bibinfo{journal}{Physical Review Letters} \textbf{\bibinfo{volume}{105}},
  \bibinfo{pages}{177002} (\bibinfo{year}{2010}).

\bibitem[{\citenamefont{Lutchyn et~al.}(2010)\citenamefont{Lutchyn, Sau, and
  {Das Sarma}}}]{Lutchyn2010}
\bibinfo{author}{\bibfnamefont{R.~M.} \bibnamefont{Lutchyn}},
  \bibinfo{author}{\bibfnamefont{J.~D.} \bibnamefont{Sau}}, \bibnamefont{and}
  \bibinfo{author}{\bibfnamefont{S.}~\bibnamefont{{Das Sarma}}},
  \bibinfo{journal}{Physical Review Letters} \textbf{\bibinfo{volume}{105}},
  \bibinfo{pages}{077001} (\bibinfo{year}{2010}).

\bibitem[{\citenamefont{Sau et~al.}(2010)\citenamefont{Sau, Lutchyn, Tewari,
  and {Das Sarma}}}]{Sau2010}
\bibinfo{author}{\bibfnamefont{J.~D.} \bibnamefont{Sau}},
  \bibinfo{author}{\bibfnamefont{R.~M.} \bibnamefont{Lutchyn}},
  \bibinfo{author}{\bibfnamefont{S.}~\bibnamefont{Tewari}}, \bibnamefont{and}
  \bibinfo{author}{\bibfnamefont{S.}~\bibnamefont{{Das Sarma}}},
  \bibinfo{journal}{Physical Review Letters} \textbf{\bibinfo{volume}{104}},
  \bibinfo{pages}{040502} (\bibinfo{year}{2010}).

\bibitem[{\citenamefont{Deng et~al.}(2012)\citenamefont{Deng, Yu, Huang,
  Larsson, Caroff, and Xu}}]{Deng2012}
\bibinfo{author}{\bibfnamefont{M.~T.} \bibnamefont{Deng}},
  \bibinfo{author}{\bibfnamefont{C.~L.} \bibnamefont{Yu}},
  \bibinfo{author}{\bibfnamefont{G.~Y.} \bibnamefont{Huang}},
  \bibinfo{author}{\bibfnamefont{M.}~\bibnamefont{Larsson}},
  \bibinfo{author}{\bibfnamefont{P.}~\bibnamefont{Caroff}}, \bibnamefont{and}
  \bibinfo{author}{\bibfnamefont{H.~Q.} \bibnamefont{Xu}},
  \bibinfo{journal}{Nano Letters} \textbf{\bibinfo{volume}{12}},
  \bibinfo{pages}{6414} (\bibinfo{year}{2012}).

\bibitem[{\citenamefont{Rokhinson et~al.}(2012)\citenamefont{Rokhinson, Liu,
  and Furdyna}}]{Rokhinson2012}
\bibinfo{author}{\bibfnamefont{L.~P.} \bibnamefont{Rokhinson}},
  \bibinfo{author}{\bibfnamefont{X.}~\bibnamefont{Liu}}, \bibnamefont{and}
  \bibinfo{author}{\bibfnamefont{J.~K.} \bibnamefont{Furdyna}},
  \bibinfo{journal}{Nature Physics} \textbf{\bibinfo{volume}{8}},
  \bibinfo{pages}{795} (\bibinfo{year}{2012}).

\bibitem[{\citenamefont{Mourik et~al.}(2012)\citenamefont{Mourik, Zuo, Frolov,
  Plissard, Bakkers, and Kouwenhoven}}]{Mourik2012}
\bibinfo{author}{\bibfnamefont{V.}~\bibnamefont{Mourik}},
  \bibinfo{author}{\bibfnamefont{K.}~\bibnamefont{Zuo}},
  \bibinfo{author}{\bibfnamefont{S.~M.} \bibnamefont{Frolov}},
  \bibinfo{author}{\bibfnamefont{S.~R.} \bibnamefont{Plissard}},
  \bibinfo{author}{\bibfnamefont{E.~P. A.~M.} \bibnamefont{Bakkers}},
  \bibnamefont{and} \bibinfo{author}{\bibfnamefont{L.~P.}
  \bibnamefont{Kouwenhoven}}, \bibinfo{journal}{Science (New York, N.Y.)}
  \textbf{\bibinfo{volume}{336}}, \bibinfo{pages}{1003} (\bibinfo{year}{2012}).

\bibitem[{\citenamefont{Vaitiek{\.e}nas
  et~al.}(2018)\citenamefont{Vaitiek{\.e}nas, Deng, Krogstrup, and
  Marcus}}]{Vaitiekenas2018}
\bibinfo{author}{\bibfnamefont{S.}~\bibnamefont{Vaitiek{\.e}nas}},
  \bibinfo{author}{\bibfnamefont{M.-T.} \bibnamefont{Deng}},
  \bibinfo{author}{\bibfnamefont{P.}~\bibnamefont{Krogstrup}},
  \bibnamefont{and} \bibinfo{author}{\bibfnamefont{C.}~\bibnamefont{Marcus}},
  \bibinfo{journal}{arXiv preprint arXiv:1809.05513}  (\bibinfo{year}{2018}).

\bibitem[{\citenamefont{Albrecht et~al.}(2016)\citenamefont{Albrecht,
  Higginbotham, Madsen, Kuemmeth, Jespersen, Nyg{\aa}rd, Krogstrup, and
  Marcus}}]{Albrecht2016}
\bibinfo{author}{\bibfnamefont{S.~M.} \bibnamefont{Albrecht}},
  \bibinfo{author}{\bibfnamefont{A.~P.} \bibnamefont{Higginbotham}},
  \bibinfo{author}{\bibfnamefont{M.}~\bibnamefont{Madsen}},
  \bibinfo{author}{\bibfnamefont{F.}~\bibnamefont{Kuemmeth}},
  \bibinfo{author}{\bibfnamefont{T.~S.} \bibnamefont{Jespersen}},
  \bibinfo{author}{\bibfnamefont{J.}~\bibnamefont{Nyg{\aa}rd}},
  \bibinfo{author}{\bibfnamefont{P.}~\bibnamefont{Krogstrup}},
  \bibnamefont{and} \bibinfo{author}{\bibfnamefont{C.~M.}
  \bibnamefont{Marcus}}, \bibinfo{journal}{Nature}
  \textbf{\bibinfo{volume}{531}}, \bibinfo{pages}{206} (\bibinfo{year}{2016}).

\bibitem[{\citenamefont{{Lutchyn} et~al.}(2018)\citenamefont{{Lutchyn},
  {Bakkers}, {Kouwenhoven}, {Krogstrup}, {Marcus}, and {Oreg}}}]{Lutchyn2018}
\bibinfo{author}{\bibfnamefont{R.~M.} \bibnamefont{{Lutchyn}}},
  \bibinfo{author}{\bibfnamefont{E.~P.~A.~M.} \bibnamefont{{Bakkers}}},
  \bibinfo{author}{\bibfnamefont{L.~P.} \bibnamefont{{Kouwenhoven}}},
  \bibinfo{author}{\bibfnamefont{P.}~\bibnamefont{{Krogstrup}}},
  \bibinfo{author}{\bibfnamefont{C.~M.} \bibnamefont{{Marcus}}},
  \bibnamefont{and} \bibinfo{author}{\bibfnamefont{Y.}~\bibnamefont{{Oreg}}},
  \bibinfo{journal}{Nature Reviews Materials} \textbf{\bibinfo{volume}{3}},
  \bibinfo{pages}{52} (\bibinfo{year}{2018}), \eprint{1707.04899}.

\bibitem[{\citenamefont{Makhlin et~al.}(2001)\citenamefont{Makhlin, Sch\"on,
  and Shnirman}}]{Schon2001}
\bibinfo{author}{\bibfnamefont{Y.}~\bibnamefont{Makhlin}},
  \bibinfo{author}{\bibfnamefont{G.}~\bibnamefont{Sch\"on}}, \bibnamefont{and}
  \bibinfo{author}{\bibfnamefont{A.}~\bibnamefont{Shnirman}},
  \bibinfo{journal}{Rev. Mod. Phys.} \textbf{\bibinfo{volume}{73}},
  \bibinfo{pages}{357} (\bibinfo{year}{2001}).

\bibitem[{\citenamefont{Glazman and Larkin}(1997)}]{Glazman1997}
\bibinfo{author}{\bibfnamefont{L.~I.} \bibnamefont{Glazman}} \bibnamefont{and}
  \bibinfo{author}{\bibfnamefont{A.~I.} \bibnamefont{Larkin}},
  \bibinfo{journal}{Phys. Rev. Lett.} \textbf{\bibinfo{volume}{79}},
  \bibinfo{pages}{3736} (\bibinfo{year}{1997}).

\bibitem[{\citenamefont{Barkeshli and Sau}(2015)}]{Barkeshli2015}
\bibinfo{author}{\bibfnamefont{M.}~\bibnamefont{Barkeshli}} \bibnamefont{and}
  \bibinfo{author}{\bibfnamefont{J.~D.} \bibnamefont{Sau}}
  (\bibinfo{year}{2015}), \eprint{1509.07135}.

\bibitem[{\citenamefont{Plugge et~al.}(2016)\citenamefont{Plugge, Landau, Sela,
  Altland, Flensberg, and Egger}}]{Plugge2016}
\bibinfo{author}{\bibfnamefont{S.}~\bibnamefont{Plugge}},
  \bibinfo{author}{\bibfnamefont{L.~A.} \bibnamefont{Landau}},
  \bibinfo{author}{\bibfnamefont{E.}~\bibnamefont{Sela}},
  \bibinfo{author}{\bibfnamefont{A.}~\bibnamefont{Altland}},
  \bibinfo{author}{\bibfnamefont{K.}~\bibnamefont{Flensberg}},
  \bibnamefont{and} \bibinfo{author}{\bibfnamefont{R.}~\bibnamefont{Egger}},
  \bibinfo{journal}{Phys. Rev. B} \textbf{\bibinfo{volume}{94}},
  \bibinfo{pages}{174514} (\bibinfo{year}{2016}).

\bibitem[{\citenamefont{Hassler and Schuricht}(2012)}]{hassler2012strongly}
\bibinfo{author}{\bibfnamefont{F.}~\bibnamefont{Hassler}} \bibnamefont{and}
  \bibinfo{author}{\bibfnamefont{D.}~\bibnamefont{Schuricht}},
  \bibinfo{journal}{New Journal of Physics} \textbf{\bibinfo{volume}{14}},
  \bibinfo{pages}{125018} (\bibinfo{year}{2012}).

\bibitem[{\citenamefont{Plugge et~al.}(2017)\citenamefont{Plugge, Rasmussen,
  Egger, and Flensberg}}]{plugge2017majorana}
\bibinfo{author}{\bibfnamefont{S.}~\bibnamefont{Plugge}},
  \bibinfo{author}{\bibfnamefont{A.}~\bibnamefont{Rasmussen}},
  \bibinfo{author}{\bibfnamefont{R.}~\bibnamefont{Egger}}, \bibnamefont{and}
  \bibinfo{author}{\bibfnamefont{K.}~\bibnamefont{Flensberg}},
  \bibinfo{journal}{New Journal of Physics} \textbf{\bibinfo{volume}{19}},
  \bibinfo{pages}{012001} (\bibinfo{year}{2017}).

\bibitem[{\citenamefont{Karzig et~al.}(2017)\citenamefont{Karzig, Knapp,
  Lutchyn, Bonderson, Hastings, Nayak, Alicea, Flensberg, Plugge, Oreg
  et~al.}}]{karzig2017scalable}
\bibinfo{author}{\bibfnamefont{T.}~\bibnamefont{Karzig}},
  \bibinfo{author}{\bibfnamefont{C.}~\bibnamefont{Knapp}},
  \bibinfo{author}{\bibfnamefont{R.~M.} \bibnamefont{Lutchyn}},
  \bibinfo{author}{\bibfnamefont{P.}~\bibnamefont{Bonderson}},
  \bibinfo{author}{\bibfnamefont{M.~B.} \bibnamefont{Hastings}},
  \bibinfo{author}{\bibfnamefont{C.}~\bibnamefont{Nayak}},
  \bibinfo{author}{\bibfnamefont{J.}~\bibnamefont{Alicea}},
  \bibinfo{author}{\bibfnamefont{K.}~\bibnamefont{Flensberg}},
  \bibinfo{author}{\bibfnamefont{S.}~\bibnamefont{Plugge}},
  \bibinfo{author}{\bibfnamefont{Y.}~\bibnamefont{Oreg}}, \bibnamefont{et~al.},
  \bibinfo{journal}{Physical Review B} \textbf{\bibinfo{volume}{95}},
  \bibinfo{pages}{235305} (\bibinfo{year}{2017}).

\bibitem[{\citenamefont{Sagi et~al.}(2019)\citenamefont{Sagi, Ebisu, Tanaka,
  Stern, and Oreg}}]{spinliquid}
\bibinfo{author}{\bibfnamefont{E.}~\bibnamefont{Sagi}},
  \bibinfo{author}{\bibfnamefont{H.}~\bibnamefont{Ebisu}},
  \bibinfo{author}{\bibfnamefont{Y.}~\bibnamefont{Tanaka}},
  \bibinfo{author}{\bibfnamefont{A.}~\bibnamefont{Stern}}, \bibnamefont{and}
  \bibinfo{author}{\bibfnamefont{Y.}~\bibnamefont{Oreg}},
  \bibinfo{journal}{Phys. Rev. B} \textbf{\bibinfo{volume}{99}},
  \bibinfo{pages}{075107} (\bibinfo{year}{2019}).

\bibitem[{\citenamefont{Kitaev}(2006)}]{Kitaev2006}
\bibinfo{author}{\bibfnamefont{A.}~\bibnamefont{Kitaev}},
  \bibinfo{journal}{Annals of Physics} \textbf{\bibinfo{volume}{321}},
  \bibinfo{pages}{2} (\bibinfo{year}{2006}).

\bibitem[{\citenamefont{Yao and Kivelson}(2007)}]{Yao2007}
\bibinfo{author}{\bibfnamefont{H.}~\bibnamefont{Yao}} \bibnamefont{and}
  \bibinfo{author}{\bibfnamefont{S.~A.} \bibnamefont{Kivelson}},
  \bibinfo{journal}{Physical Review Letters} \textbf{\bibinfo{volume}{99}},
  \bibinfo{pages}{247203} (\bibinfo{year}{2007}).

\bibitem[{\citenamefont{Belavin et~al.}(1984)\citenamefont{Belavin, Polyakov,
  and Zamolodchikov}}]{Belavin1984}
\bibinfo{author}{\bibfnamefont{A.}~\bibnamefont{Belavin}},
  \bibinfo{author}{\bibfnamefont{A.}~\bibnamefont{Polyakov}}, \bibnamefont{and}
  \bibinfo{author}{\bibfnamefont{A.}~\bibnamefont{Zamolodchikov}},
  \bibinfo{journal}{Nuclear Physics B} \textbf{\bibinfo{volume}{241}},
  \bibinfo{pages}{333} (\bibinfo{year}{1984}).

\bibitem[{\citenamefont{Blume}(1966)}]{Blume}
\bibinfo{author}{\bibfnamefont{M.}~\bibnamefont{Blume}},
  \bibinfo{journal}{Phys. Rev.} \textbf{\bibinfo{volume}{141}},
  \bibinfo{pages}{517} (\bibinfo{year}{1966}).

\bibitem[{\citenamefont{Capel}(1966)}]{Capel}
\bibinfo{author}{\bibfnamefont{H.}~\bibnamefont{Capel}},
  \bibinfo{journal}{Physica} \textbf{\bibinfo{volume}{32}}, \bibinfo{pages}{966
  } (\bibinfo{year}{1966}).

\bibitem[{\citenamefont{Mussardo}(2017)}]{Mussardo2017}
\bibinfo{author}{\bibfnamefont{G.}~\bibnamefont{Mussardo}},
  \emph{\bibinfo{title}{Statistical Field Theory}} (\bibinfo{publisher}{Oxford
  univeristy press}, \bibinfo{year}{2017}), ISBN
  \bibinfo{isbn}{978-0199547586}.

\bibitem[{\citenamefont{Zamolodchikov}(1986)}]{Zamolodchikov1986}
\bibinfo{author}{\bibfnamefont{A.~B.} \bibnamefont{Zamolodchikov}},
  \bibinfo{journal}{Sov. J. Nucl. Phys.} \textbf{\bibinfo{volume}{44}},
  \bibinfo{pages}{529} (\bibinfo{year}{1986}).

\bibitem[{\citenamefont{Kastor et~al.}(1989)\citenamefont{Kastor, Martinec, and
  Shenker}}]{Kastor1989}
\bibinfo{author}{\bibfnamefont{D.}~\bibnamefont{Kastor}},
  \bibinfo{author}{\bibfnamefont{E.}~\bibnamefont{Martinec}}, \bibnamefont{and}
  \bibinfo{author}{\bibfnamefont{S.}~\bibnamefont{Shenker}},
  \bibinfo{journal}{Nuclear Physics B} \textbf{\bibinfo{volume}{316}},
  \bibinfo{pages}{590} (\bibinfo{year}{1989}).

\bibitem[{\citenamefont{Rahmani et~al.}(2015)\citenamefont{Rahmani, Zhu, Franz,
  and Affleck}}]{Affleck2015}
\bibinfo{author}{\bibfnamefont{A.}~\bibnamefont{Rahmani}},
  \bibinfo{author}{\bibfnamefont{X.}~\bibnamefont{Zhu}},
  \bibinfo{author}{\bibfnamefont{M.}~\bibnamefont{Franz}}, \bibnamefont{and}
  \bibinfo{author}{\bibfnamefont{I.}~\bibnamefont{Affleck}},
  \bibinfo{journal}{Phys. Rev. Lett.} \textbf{\bibinfo{volume}{115}},
  \bibinfo{pages}{166401} (\bibinfo{year}{2015}).

\bibitem[{\citenamefont{Zhu and Franz}(2016)}]{Franz2016}
\bibinfo{author}{\bibfnamefont{X.}~\bibnamefont{Zhu}} \bibnamefont{and}
  \bibinfo{author}{\bibfnamefont{M.}~\bibnamefont{Franz}},
  \bibinfo{journal}{Phys. Rev. B} \textbf{\bibinfo{volume}{93}},
  \bibinfo{pages}{195118} (\bibinfo{year}{2016}).

\bibitem[{\citenamefont{Grover et~al.}(2014)\citenamefont{Grover, Sheng, and
  Vishwanath}}]{Grover2014}
\bibinfo{author}{\bibfnamefont{T.}~\bibnamefont{Grover}},
  \bibinfo{author}{\bibfnamefont{D.}~\bibnamefont{Sheng}}, \bibnamefont{and}
  \bibinfo{author}{\bibfnamefont{A.}~\bibnamefont{Vishwanath}},
  \bibinfo{journal}{Science} \textbf{\bibinfo{volume}{344}},
  \bibinfo{pages}{280} (\bibinfo{year}{2014}).

\bibitem[{\citenamefont{O’Brien and Fendley}(2018)}]{o2018lattice}
\bibinfo{author}{\bibfnamefont{E.}~\bibnamefont{O’Brien}} \bibnamefont{and}
  \bibinfo{author}{\bibfnamefont{P.}~\bibnamefont{Fendley}},
  \bibinfo{journal}{Physical review letters} \textbf{\bibinfo{volume}{120}},
  \bibinfo{pages}{206403} (\bibinfo{year}{2018}).

\bibitem[{\citenamefont{Fradkin and Susskind}(1978)}]{Fradkin1978}
\bibinfo{author}{\bibfnamefont{E.}~\bibnamefont{Fradkin}} \bibnamefont{and}
  \bibinfo{author}{\bibfnamefont{L.}~\bibnamefont{Susskind}},
  \bibinfo{journal}{Phys. Rev. D} \textbf{\bibinfo{volume}{17}},
  \bibinfo{pages}{2637} (\bibinfo{year}{1978}).

\bibitem[{\citenamefont{Friedan et~al.}(1985)\citenamefont{Friedan, Qiu, and
  Shenker}}]{Friedan}
\bibinfo{author}{\bibfnamefont{D.}~\bibnamefont{Friedan}},
  \bibinfo{author}{\bibfnamefont{Z.}~\bibnamefont{Qiu}}, \bibnamefont{and}
  \bibinfo{author}{\bibfnamefont{S.}~\bibnamefont{Shenker}},
  \bibinfo{journal}{Physics Letters B} \textbf{\bibinfo{volume}{151}},
  \bibinfo{pages}{37 } (\bibinfo{year}{1985}).

\bibitem[{\citenamefont{Saul et~al.}(1974)\citenamefont{Saul, Wortis, and
  Stauffer}}]{Saul}
\bibinfo{author}{\bibfnamefont{D.~M.} \bibnamefont{Saul}},
  \bibinfo{author}{\bibfnamefont{M.}~\bibnamefont{Wortis}}, \bibnamefont{and}
  \bibinfo{author}{\bibfnamefont{D.}~\bibnamefont{Stauffer}},
  \bibinfo{journal}{Phys. Rev. B} \textbf{\bibinfo{volume}{9}},
  \bibinfo{pages}{4964} (\bibinfo{year}{1974}).

\bibitem[{\citenamefont{von Gehlen}(1990)}]{Gehlen}
\bibinfo{author}{\bibfnamefont{G.}~\bibnamefont{von Gehlen}},
  \bibinfo{journal}{Nuclear Physics B} \textbf{\bibinfo{volume}{330}},
  \bibinfo{pages}{741 } (\bibinfo{year}{1990}).

\bibitem[{spl()}]{spl}
\bibinfo{note}{See supplementary material, which includes Refs.
  \cite{Villain,Wess}.}

\bibitem[{\citenamefont{Schrieffer and Wolff}(1966)}]{Wolf}
\bibinfo{author}{\bibfnamefont{J.~R.} \bibnamefont{Schrieffer}}
  \bibnamefont{and} \bibinfo{author}{\bibfnamefont{P.~A.} \bibnamefont{Wolff}},
  \bibinfo{journal}{Phys. Rev.} \textbf{\bibinfo{volume}{149}},
  \bibinfo{pages}{491} (\bibinfo{year}{1966}).

\bibitem[{\citenamefont{Gogolin et~al.}(2004)\citenamefont{Gogolin, Nersesyan,
  and Tsvelik}}]{Gogolin}
\bibinfo{author}{\bibfnamefont{A.}~\bibnamefont{Gogolin}},
  \bibinfo{author}{\bibfnamefont{A.}~\bibnamefont{Nersesyan}},
  \bibnamefont{and} \bibinfo{author}{\bibfnamefont{A.}~\bibnamefont{Tsvelik}},
  \emph{\bibinfo{title}{Bosonization and Strongly Correlated Systems}}
  (\bibinfo{publisher}{Cambridge University Press}, \bibinfo{year}{2004}), ISBN
  \bibinfo{isbn}{9780521617192}.

\bibitem[{foo({\natexlab{a}})}]{foot1}
\bibinfo{note}{Compared to Eq.~(10) where $t$ is in units of energy, here $t$
  has a dimension of velocity and is formally scaled by the lattice constant
  (the typical distance between the MZMs). For simplicity, we omit this scaling
  factor. If one writes this factor explicitly, then $t$ is modified to $tl_0$
  ($l_0$ is lattice spacing) and taking the limit of $l_0\to 0$ while
  $\tilde{t}=tl_0$ is fixed. Renaming $\tilde{t}$ as $t$, we get Eq.~(12)}.

\bibitem[{\citenamefont{Vecchia and Ferrara}(1977)}]{DIVECCHIA}
\bibinfo{author}{\bibfnamefont{P.~D.} \bibnamefont{Vecchia}} \bibnamefont{and}
  \bibinfo{author}{\bibfnamefont{S.}~\bibnamefont{Ferrara}},
  \bibinfo{journal}{Nuclear Physics B} \textbf{\bibinfo{volume}{130}},
  \bibinfo{pages}{93 } (\bibinfo{year}{1977}).

\bibitem[{foo({\natexlab{b}})}]{foot3}
\bibinfo{note}{The definition of $\theta, \bar \theta$, $D$, $F$ and the
  equivalence between the models in Eqs.~(18), (19), and (20) are provided in
  the Supplemental Material.}

\bibitem[{\citenamefont{Witten}(1978)}]{Witten}
\bibinfo{author}{\bibfnamefont{E.}~\bibnamefont{Witten}},
  \bibinfo{journal}{Nuclear Physics B} \textbf{\bibinfo{volume}{142}},
  \bibinfo{pages}{285 } (\bibinfo{year}{1978}).

\bibitem[{\citenamefont{Christe and Mussardo}(1990)}]{Christe1990}
\bibinfo{author}{\bibfnamefont{P.}~\bibnamefont{Christe}} \bibnamefont{and}
  \bibinfo{author}{\bibfnamefont{G.}~\bibnamefont{Mussardo}},
  \bibinfo{journal}{Nuclear Physics B} \textbf{\bibinfo{volume}{330}},
  \bibinfo{pages}{465} (\bibinfo{year}{1990}).

\bibitem[{\citenamefont{Kane and Fisher}(1992)}]{KaneFisher}
\bibinfo{author}{\bibfnamefont{C.~L.} \bibnamefont{Kane}} \bibnamefont{and}
  \bibinfo{author}{\bibfnamefont{M.~P.~A.} \bibnamefont{Fisher}},
  \bibinfo{journal}{Phys. Rev. B} \textbf{\bibinfo{volume}{46}},
  \bibinfo{pages}{15233} (\bibinfo{year}{1992}).

\bibitem[{\citenamefont{Banerjee et~al.}(2018)\citenamefont{Banerjee, Heiblum,
  Umansky, Feldman, Oreg, and Stern}}]{banerjee2018observation}
\bibinfo{author}{\bibfnamefont{M.}~\bibnamefont{Banerjee}},
  \bibinfo{author}{\bibfnamefont{M.}~\bibnamefont{Heiblum}},
  \bibinfo{author}{\bibfnamefont{V.}~\bibnamefont{Umansky}},
  \bibinfo{author}{\bibfnamefont{D.~E.} \bibnamefont{Feldman}},
  \bibinfo{author}{\bibfnamefont{Y.}~\bibnamefont{Oreg}}, \bibnamefont{and}
  \bibinfo{author}{\bibfnamefont{A.}~\bibnamefont{Stern}},
  \bibinfo{journal}{Nature} p.~\bibinfo{pages}{1} (\bibinfo{year}{2018}).

\bibitem[{\citenamefont{Villain}(1975)}]{Villain}
\bibinfo{author}{\bibfnamefont{J.}~\bibnamefont{Villain}},
  \bibinfo{journal}{Journal de Physique} \textbf{\bibinfo{volume}{36}},
  \bibinfo{pages}{581} (\bibinfo{year}{1975}).

\bibitem[{\citenamefont{Wess and Bagger}(1992)}]{Wess}
\bibinfo{author}{\bibfnamefont{J.}~\bibnamefont{Wess}} \bibnamefont{and}
  \bibinfo{author}{\bibfnamefont{J.~A.} \bibnamefont{Bagger}},
  \emph{\bibinfo{title}{{Supersymmetry and supergravity; 2nd ed.}}}, Princeton
  Series in Physics (\bibinfo{publisher}{Princeton Univ. Press},
  \bibinfo{address}{Princeton, NJ}, \bibinfo{year}{1992}).

\end{thebibliography}




\end{document}